\documentclass[manuscript]{acmart}

\AtBeginDocument{%
  \providecommand\BibTeX{{%
    \normalfont B\kern-0.5em{\scshape i\kern-0.25em b}\kern-0.8em\TeX}}}

\usepackage[ruled]{algorithm2e}
\usepackage{xpatch}
\usepackage{graphicx, subcaption}

\begin{document}

\settopmatter{printacmref=false} 
\renewcommand\footnotetextcopyrightpermission[1]{} 
\pagestyle{plain} 
\setcopyright{none}
\makeatletter
\xpatchcmd{\ps@firstpagestyle}{Manuscript submitted to ACM}{}{\typeout{First patch succeeded}}{\typeout{first patch failed}}
\xpatchcmd{\ps@standardpagestyle}{Manuscript submitted to ACM}{}{\typeout{Second patch succeeded}}{\typeout{Second patch failed}}    \@ACM@manuscriptfalse
\makeatother
\title{MagSurface: Wireless 2D Finger Tracking Leveraging Magnetic Fields}


\author{Sarnab Bhattacharya}
\email{sarnab2008@utexas.edu}
\affiliation{%
  \institution{University of Texas at Austin}
  \streetaddress{2501 Speedway}
  \city{Austin}
  \state{Texas}
  \postcode{78705}
}

\author{Keum San Chun}
\email{gmountk@gmail.com}
\affiliation{%
  \institution{University of Texas at Austin}
  \streetaddress{2501 Speedway}
  \city{Austin}
  \state{Texas}
  \postcode{78705}
}

\author{Edison Thomaz}
\email{ethomaz@utexas.edu}
\affiliation{%
  \institution{University of Texas at Austin}
  \streetaddress{2501 Speedway}
  \city{Austin}
  \state{Texas}
  \postcode{78705}
}


\renewcommand{\shortauthors}{Sarnab, et al.}

\begin{abstract}
  With the ubiquity of touchscreens, touch input modality has become a popular way of interaction. However, current touchscreen technology is limiting in its design as it restricts touch interactions to specially instrumented touch surfaces. Surface contaminants like water can also hinder proper interactions. In this paper, we propose the use of magnetic field sensing to enable finger tracking on a surface with minimal instrumentation. Our system, MagSurface, turns everyday surfaces into a touch medium, thus allowing more flexibility in the types of touch surfaces. The evaluation of our system consists of quantifying the accuracy of the system in locating an object on 2D flat surfaces. We test our system on three different surface materials to validate its usage scenarios. A qualitative user experience study was also conducted to get feedback on the ease of use and comfort of the system. Localization error as low as a few millimeters was achieved.
\end{abstract}




\begin{teaserfigure}
    \centering
    \begin{subfigure}[b]{0.59\linewidth}
        \includegraphics[width=1\linewidth]{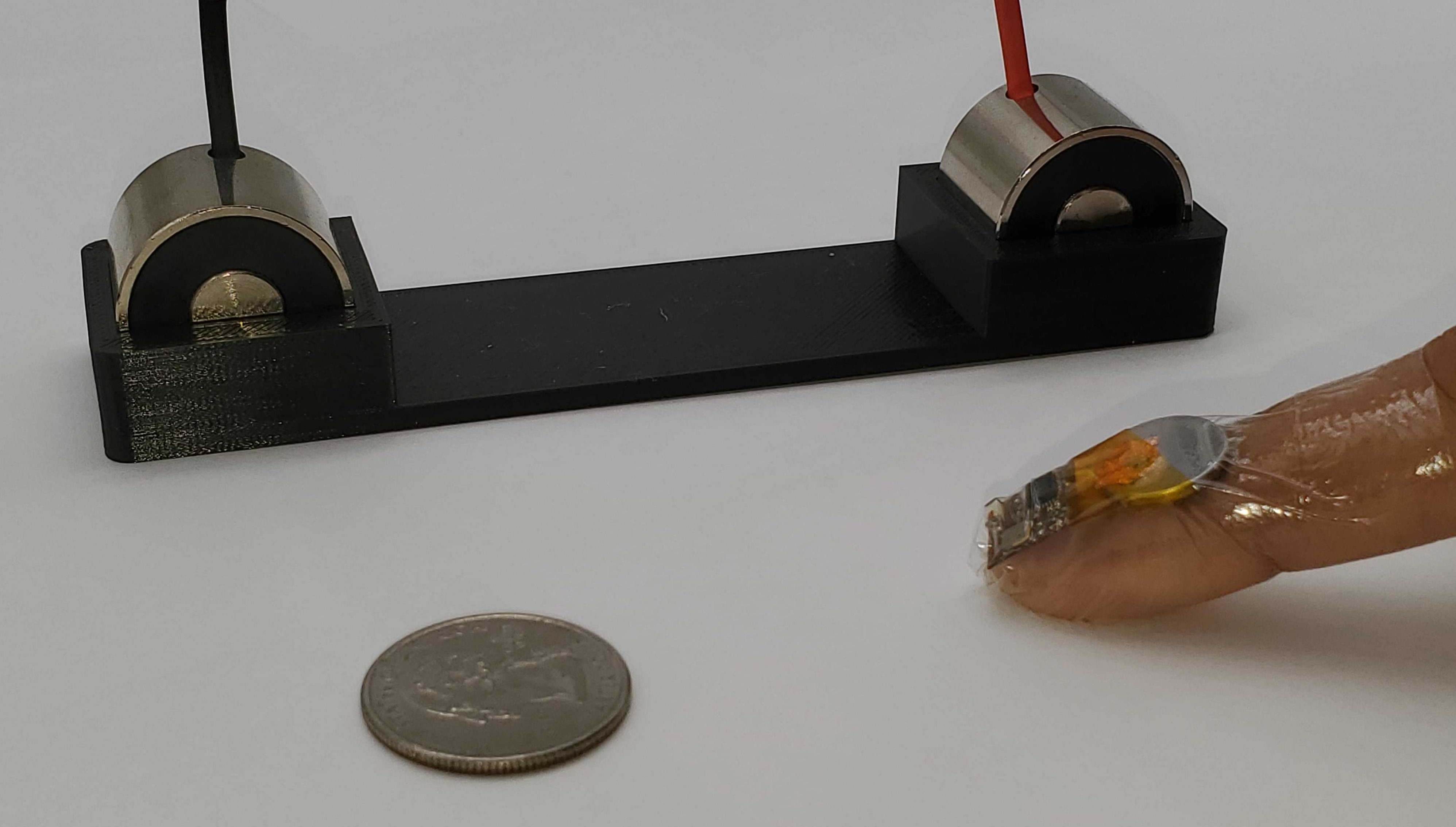}
        \caption{\label{fig:SystemComp}}
    \end{subfigure}
    \begin{subfigure}[b]{0.4\linewidth}
        \includegraphics[width=1\linewidth]{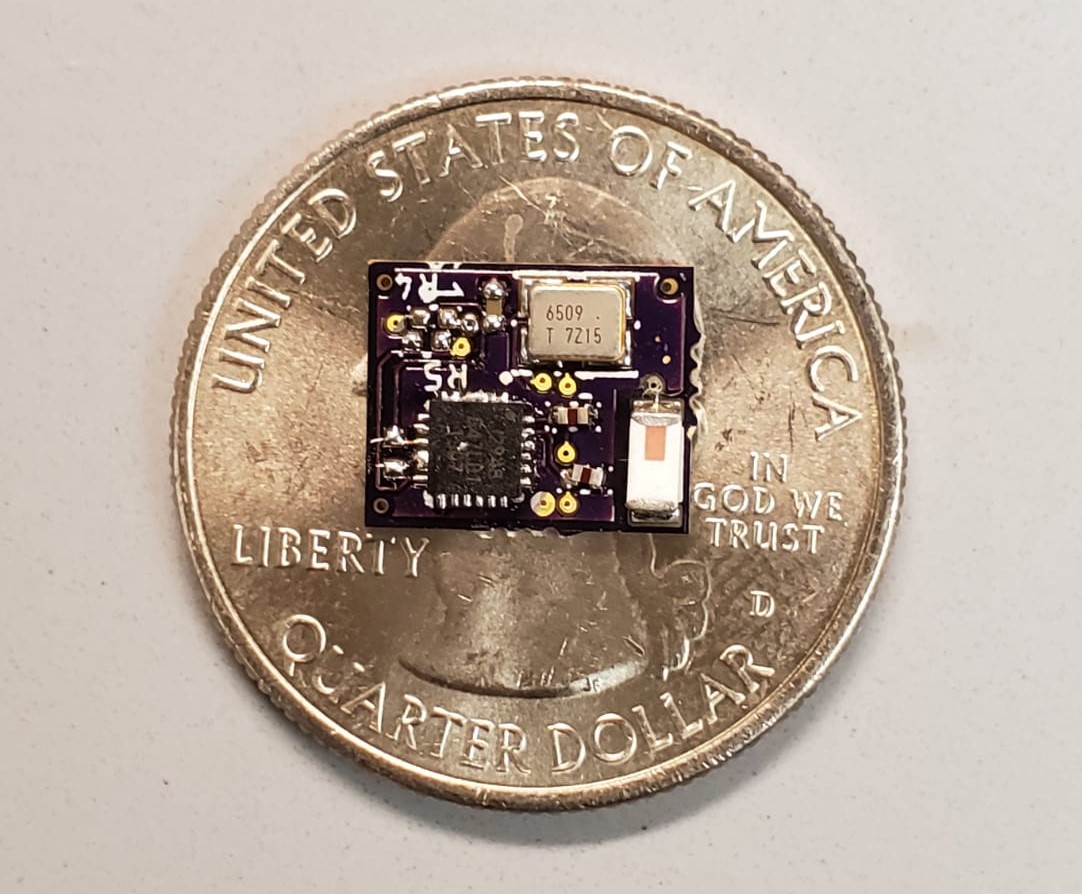}
        \caption{\label{fig:device_img}}
    \end{subfigure}
    \caption{(a) MagSurface System consisting of electromagnets mounted on a custom designed 3D printed structure and wearable band-aid, and, (b) The wireless finger tracking device was built on a flexible PCB and consists of three functional blocks: wireless micro-controller unit (MCU), sensing unit, and antenna.}
\end{teaserfigure}

\maketitle

\section{Introduction}
Over the last decade, touch input has become the most common human-computer interface for interacting with mobile devices and large displays. The popularity of touchscreens, and the intuitive manner in which they support direct manipulation \cite{hutchins1985direct} has brought touch technologies to a variety of physical interfaces that require user input such as computer screens, music players, ebook readers and automotive dashboards. However, while largely commoditized for smartphones and personal devices, current touchscreen technology is difficult to adapt for every applications that require human machine interaction. This is because today's touchscreens are most commonly built using resistive or capacitive sensing technology applied to a glass surface. As a result, precision touchscreens tend to be not only relatively small and expensive, but also suffer from a few important shortcomings, such as not performing well in the presence of water, e.g., in the rain, and being available only as a strictly smooth surface. In light of these drawbacks, the research community has explored a variety of techniques to enhance touch based human machine interaction paradigms. These have included supplementing objects with touch input modality, and allowing for more flexibility in the types of surfaces where touch technologies can be applied \cite{ono2013touch,poupyrev2016project,Sato:2012:TET:2207676.2207743,Xiao:2013:WRE:2470654.2466113}.

In this paper, we present a technique for wireless finger tracking based on magnetic field sensing. Our approach extends touch input to any flat surface, addressing some of the limitations of current touchscreen technologies. We demonstrate our technique through a system implementation called MagSurface. MagSurface is comprised of two components, (1) a pair of electromagnets and (2) a custom wearable device, both shown in figure \ref{fig:SystemComp}. The electromagnets generate a time varying magnetic field which is captured by a magnetometer in the wearable device, which is small and can be easily attached to the user's fingertips or place on a ring. Limitations of capacitive touchscreens such as sub-optimal operation in the presence of water or dirt do not hinder our approach. Likewise, our method is not affected by irregularities of the tracking surface, i.e., the tracking surface does not have to be smooth. These attributes make our approach particularly suitable for applications situated in outdoor environments (e.g., industrial applications), in settings where there might be a need to operate a computing device with dirty hands or through a dirty surface, (e.g., kitchen, machine shop), and in environments whose input surfaces must be sterilized with chemicals that could impact or damage the electronics in traditional touchscreens.

The contributions of this work can be summarized as follows:

\begin{itemize}
    \item A system design and implementation comprised of an electromagnet bar for generating magnetic fields and a lightweight, low power, and wireless wearable to perform tracking of a user's fingers on 2D surfaces.
    \item An algorithm for finger tracking by iteratively reducing position error which can be run on an embedded micro-controller.
    \item An evaluation of the system on 3 surface materials, metallic, wood and acrylic, in which we showed our approach produces localization errors not greater than 0.25cm.
    \item A qualitative user experience survey with 12 participants, whose results demonstrated the viability of the approach and highlighted areas for improvement.
\end{itemize}

\section{Related Work}
Before touchscreens in mobile devices, finger tracking was made popular through trackpads on laptop computers, which have been long studied \cite{epps1986comparison}, and custom camera-based multi-touch systems such as Microsoft Surface \cite{de2010function}. Over the years, various technologies have been developed to support touch including Interpolated Force Sensitive Resistance (IFSR), which support multitouch interactions and can acquire high-quality anti-aliased pressure images at high frame rates \cite{Rosenberg:2009:UIM:1576246.1531371}. 

Current touchscreens typically use transparent resistive or capacitive panels overlayed on glass screens. However, as touch interactions have become increasingly more popular, a large body of research has examined new technical approaches for enabling touch-enabled interfaces in everyday surfaces and objects. These methods have explored a wide range of sensing modalities and form factors. For instance, Takeoka et al. installed multi-layered infrared lasers to detect not only hand and finger contact but also their orientation \cite{Takeoka:2010:ZIG:1936652.1936668}. SmartSkin instrumented an entire tabletop with capacitive leads to act as a touch interface, with support for hover detection \cite{Rekimoto:2002:SIF:503376.503397}. The sections below describe several techniques that have been proposed by researchers towards improving and augmenting touch interfaces.
    
    \subsection{Optical Sensing}
    A large body of work focusing on vision-based and optical-based finger tracking exist. Harrison et al. and others experimented with a portable projector and Kinect system to enable touch and gesture interactions anywhere \cite{Harrison:2011:OWM:2047196.2047255}. Similar methods include \cite{Wilson:2005:PCI:1095034.1095047}\cite{Xiao:2013:WRE:2470654.2466113}\cite{Wilson:2004:TIT:1027933.1027946}. SymmetriSense is a system utilizing commodity cameras to detect touches on glossy surfaces \cite{Yoo:2016:SEN:2858036.2858286}. Frustrated Total Internal Reflection (FTIR) methods, which can be thought of as a combination of optical and vision based system like \cite{Hofer:2009:FFM:1517664.1517730}\cite{Echtler:2009:IFE:1731903.1731909}, have also yielded good results. Commercial products like EvoMouse \cite{evomouse} and LeapMotion \cite{leap} have used Infra Red (IR) reflection based sensing to track user fingers.
    
    \subsection{Acoustic Sensing}
    Acoustic approaches to detect touches, taps, knocks or flicks on surfaces have been extensively explored. Kim et al. used commercial smartphone microphones to enable tap interactions on a variety of surfaces \cite{Kim:2018:ULA:3274783.3274848}. Other sound-based techniques to localize touch events and gestures have also been proposed by \cite{Sun:2018:VST:3241539.3241568}\cite{7964907}\cite{5655110}. Acoustic approaches can be efficient in detecting and localizing taps. However, tracing a finger as it is dragged on a surface can be more challenging with a sound-only approach.
    SoundTrak, on the other hand, performs continuous finger tracking in 3D space by capturing and processing sound generated from a small speaker attached to the user's finger \cite{Zhang:2017:SCT:3120957.3090095}.
    
    \subsection{Magnetic Field Sensing}
    Magnetic Field (MF) sensing has long been considered a viable, accurate and occlusion-free method of tracking objects. Raab et al. details a series of transformations to convert non-linear magnetic field equations to directly solvable linear equations \cite{4102227}. This sets the foundation for using MF sensing to track objects in 3D space. 
    Some simple approaches include using permanent magnets attached to fingers \cite{Harrison:2009:AWH:1622176.1622199}, or in rings \cite{Ashbrook:2011:NSE:1978942.1979238} to provide 1D input to smartwatches. Others proposed covering an entire interaction space with magnetic sensors \cite{Liang:2012:GAS:2380116.2380157}. Han et al. demonstrated that it is possible to track a magnet in 2D using a pair of magnetometers \cite{4421009}. However, their assumptions made about the relative orientations of the magnet might not always hold true in all scenarios. uTrack built on this concept to propose a system for tracking the position of a permanent magnet in 3D space \cite{Chen:2013:UIU:2501988.2502035}. However, they relied on an exhaustive search of the orientation space which may be too computation intensive for real-time low power embedded applications.
    Chen et al. proposed a system to track fingers with high precision in 3D \cite{Chen:2016:FTP:2858036.2858125}. Active electromagnets instead of permanent magnets were used to filter out stray or ambient magnetic fields. Four magnetometers were used to resolve the positioning ambiguity that needs to be overcome when using MF sensing. The magnetometers were placed on a fixed base while the active electromagnets were placed on fingers. Dai et al. used both active electromagnets and geomagnetic fields to perform 6D tracking using one electromagnet and one magnetometer \cite{dai20176}. Their algorithm relied on geomagnetic fields for orientation and used that information to calculate the 3D position. This algorithm is quite complex and would be hard to perform on a small low cost embedded processor in real time. Ge et al. proposed a unique aparatus with rotating electromagnets \cite{ge2009novel}. In theory it is a novel implementation but mechanical components like rotating electromagnets are difficult to implement and are prone to failures.
    
    \subsection{Instrumented Objects}
    Finally, there has been strong interest in sensing touch in irregular shapes and beyond surfaces. Sato et al. adopted an approach of embedding capacitive sensors on everyday objects to sense touch and gesture \cite{Sato:2012:TET:2207676.2207743}.
    Skintrack instrumented the human arm by attaching electrodes on the wrist and wearing a signal emitting ring on the finger to continuously track finger movement on the forearm and palm \cite{Zhang:2016:SUB:2858036.2858082}.
    Zhang et al. proposes a unique way of either fabricating objects, or coating objects with a conductive material \cite{Zhang:2017:ELT:3025453.3025842}. Electrodes are then attached to the object and using electric field tomography, touch and finger tracking can be performed. Instrumenting surfaces or objects provide very accurate detection and tracking of touch and other input modalities like pressure orientation, etc. But these approaches require extra hardware and computational power. They also tend to be specific to the particular setting, and thus are not easily portable.

\section{Motivation}
Most acoustic approaches suffer from being unable to continuously track objects. They are also susceptible to noise. Thus it would not be ideal to use them to simulate or replace a touchpad or touchscreen. Multi-finger interactions would also be hard to integrate using acoustic sensing.
Occlusion was a major issue with vision/optical based systems. Vision based systems also have to perform intensive computations which would not be ideal for a low power embedded device. 
MF sensing is not hindered by occlusion and addresses this issue. But in most of the previous work, some previous assumption about the relative orientation was made or a compute intensive search was conducted to find the orientations. Chen et al.\cite{Chen:2016:FTP:2858036.2858125} came up with a orientation invariant method of tracking objects in 3D using MF sensing but their system utilized a lot of hardware to surmount the problem of not knowing relative orientations apriori.

Our design was driven by the desire to avoid instrumenting the interaction space with specialized and hard to set up sensors which would reduce portability and increase costs. Instead, we aimed for a portable approach; to use the system, a user would only need to place the electromagnets on a surface of choice and attach a minimal wearable piece to a finger. For user comfort, the wearable device on the finger was made very small (10x8 \(mm^2\)) as shown in figure \ref{fig:device_img}. Bluetooth Low Energy(BLE) was chosen as the wireless link for communicating data from the system as most consumer devices come with BLE chips built-in and thus would offer easier integration. 2D tracking offers some benefits with regards to reduction in the amount of hardware needed to accurately track objects using MF sensing. The magnetometer is attached to the user's finger and the electromagnets are used as a static anchor. This offers 3 advantages:

\begin{itemize}
    \item The wearable part is small compared to a bulky electromagnet and would offer more ease of use.
    \item It is low power and thus can be powered by a small coin cell battery. The electromagnets on the surface can be powered by larger batteries and can even be built into devices or furniture.
    \item To expand the system for multi-finger use, more magnetometers would need to be used. This is easier than adding another electromagnet with its characteristic AC frequency and accordingly adding another filter on the processing side.
\end{itemize}

Exploiting the geometry of the setup and the constraint that the interaction space is 2D, the system uses minimal hardware and algorithms that are not computationally intensive to track a user's finger.

\section{Algorithm}
\begin{figure}[t]
    \centering
    \begin{subfigure}[b]{0.45\linewidth}
        \includegraphics[width=1.0\linewidth]{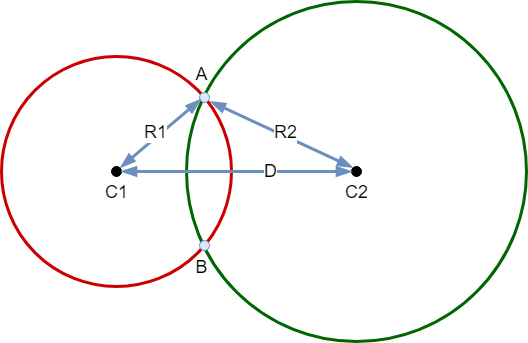}
        \begin{minipage}{0.1cm}
            \vfill
        \end{minipage}
        \caption{\label{fig:trilateration}}
    \end{subfigure}
    \hspace{0.06\linewidth}
    \begin{subfigure}[b]{0.45\linewidth}
        \includegraphics[width=1.0\linewidth]{Images/2Dmagnetism.png}
        \caption{\label{fig:fieldlines}}
    \end{subfigure}
    \caption{(a) Trilateration with 2 anchors (C1 and C2) gives 2 probable locations (A and B) of where the object might be. The distance D between the 2 anchors and the 2 radii R1 and R2 are known quantities. A is where our object actually is, and, (b) Magnetic field lines generated by a small magnetic source projected on a 2D plane. The magnetic field vector at the sensor can be decomposed into the radial and angular components}
\end{figure}

MagSurface uses inexpensive off the shelf hardware. We have designed some simple, intuitive algorithms to utilize this hardware setup to accurately track objects in 2D. Our method uses a modified trilateration technique to calculate the distance between the magnetometer and the electromagnets based on magnet field strength (figure \ref{fig:trilateration}). MagSurface can be placed on any flat surface and our current implementation is capable of 2D finger tracking within a 10x10 \(cm^2\) rectangular area.

    \subsection{Trilateration}
    Central to our tracking algorithm is trilateration, a technique that consists of being able to determine the exact location of an object if the distance of the object from anchors position points are known. Trilateration, as the name suggests, implies the need to know the distance between an object and 3 anchor points to calculate the location of an object. This is true for locating an object on a 2D surface.
    
    With 3 anchor points, it is possible to precisely calculate the location of an object in 2D space. But in our case we only have two anchors -- the two electromagnets. Using two anchors, we can localize the object to two locations (A and B) as illustrated in Figure \ref{fig:trilateration}. The center of the two circles (C1 and C2) are the two electromagnets. In our specific case we have assumed that interactions by a particular user will only be performed on one side of the electromagnets. Thus we can safely discard one of the two possible locations where our target object could be and lock in on the exact location of the tracked object.
    
    
    \subsection{Magnetic Positioning}
    Our approach leverages magnetic waves generated by alternating voltage to obtain the distance of our object from the electromagnets, i.e. anchor points. Since the system tracks interactions on a 2D surface, we only consider magnetic field lines in the plane on which the centers of the electromagnets and the object lies, i.e. the surface on which we are interacting. From Figure \ref{fig:fieldlines} it can be observed that the strength of the magnetic field \textbf{H} needs 2 factors to be defined, the radial distance \textit{r} from the magnets center to the sensor, and the angle $\mathit{ \theta }$ between the magnets north pole and the sensor. Both of these values are unknown. \textbf{H} is generally decomposed into its radial and angular terms $\mathbf{H_r}$ and $\mathbf{H_\theta}$ respectively. The values of these quantities are given by:
    
    \begin{equation}
    ||\mathbf{H_r}|| = M\ cos\theta / 2\pi r^3
    \label{eq:Hr}
    \end{equation}
    \begin{equation}
    ||\mathbf{H_\theta}|| = M\ sin\theta / 4\pi r^3
    \label{eq:Hth}
    \end{equation}
    
    Here, M is the magnetic moment of the electromagnet and relates to the permeability of the core, current through the coil, and the cross-sectional area of the coil.
    
    \begin{algorithm}[t]
    \SetAlgoNoLine
    \KwIn{$\mathbf{H_{20}},\ \mathbf{H_{30}},\ K,\ D$ }
    \KwOut{X, Y}
    $\theta_{20} \gets 0,\ \theta_{30} \gets 0$\;
    \For{$k \gets 1$ to $N$}
    {
    $r_{20}\ calculated\ from\ \mathbf{H_{20}}\ and\ \theta_{20}\ using\ equ\ \ref{eq:mag_strength}$\;
    $r_{30}\ calculated\ from\ \mathbf{H_{30}}\ and\ \theta_{30}\ using\ equ\ \ref{eq:mag_strength}$\;
    $(X,Y)\ calculated\ from\ r_{20},\ r_{30}\ and\ D$\;
    $\theta_{20} \gets arctan(X/Y)$\;
    $\theta_{30} \gets arctan((D-X)/Y)$\;
    }
    \caption{Converging to correct $\mathit{\theta}$ and \textit{r}}
    \label{alg:convergence}
    \end{algorithm}
    
    As \cite{Chen:2016:FTP:2858036.2858125} has shown, unless the relative orientation of the electromagnet and our sensor is known, it is very difficult to compute $\mathbf{H_r}$ and $\mathbf{H_\theta}$ from the data obtained from the magnetometer. The magnetometer we used is a 3-axis magnetometer, which outputs the magnetic field strength detected on each of its axes as $\mathbf{H_x}$, $\mathbf{H_y}$ and $\mathbf{H_z}$ for x-axis, y-axis and z-axis respectively. To obtain $\mathbf{H_r}$ and $\mathbf{H_\theta}$ from these values would require either knowing the orientation beforehand and applying a transform or exhaustively searching the rotation space for the global optimal rotation that transforms ($\mathbf{H_x}$, $\mathbf{H_y}$, $\mathbf{H_z}$) to ($\mathbf{H_r}$,  $\mathbf{H_\theta}$, 0) \cite{Chen:2013:UIU:2501988.2502035}. Both of these methods are not feasible as neither the relative orientation is provided apriori nor do we want to use the computational capability to do an exhaustive search of the rotation space in real-time. So instead, we calculate the total magnetic field strength given by:
    
    \begin{equation}
    ||\textbf{H}||^2 =  H_x^2 + H_y^2 + H_z^2 = K \times r^{-6} \times (3*cos^2\theta + 1)
    \label{eq:mag_strength}
    \end{equation}
    
    The K in equation \ref{eq:mag_strength} is a constant. We still have an under-constrained equation as we are trying to figure out both \textit{r} and $\mathit{ \theta }$ from only one equation. This is similar to the approach adopted by \cite{Chen:2016:FTP:2858036.2858125}, but unlike their method of using multiple magnetometers to resolve the ambiguity, we use an iterative algorithm to accurately estimate both \textit{r} and $\mathit{ \theta }$ from one equation itself by leveraging the geometric constraints of the system. It should also be noted that these equations only hold when we treat the magnet as a small dipole. In other words, the distance from the magnet should be greater than the dimension of the magnets.
    
    Our algorithm solves the problem of estimating both \textit{r} and $\mathit{ \theta }$ from one equation by taking advantage of the geometric constraints and symmetries in our system. While calculating the distance from the wearable to one electromagnet is key, it is critical to consider the entire system as a whole. The 2 electromagnets are driven by an alternating current(AC) source. One magnet is driven by a 20Hz sine wave while the other is driven by a 30Hz sine wave. As they are on different frequencies the magnetic fields do not interfere with each other. Stray magnetic fields and the earth's magnetic field, a DC field, also do not affect the generated magnetic fields unless they fall in the operating frequency band of the electromagnets. The magnetometer on the wearable device samples at 100Hz, which has a Nyquist cut off frequency of 50Hz, well above the frequency of our electromagnets. A Fast Fourier Transform(FFT) is performed on the magnetometer data and the frequency components at 20Hz and 30Hz are obtained. This data is fed into our algorithm as $\mathbf{H_{20}}$ and $\mathbf{H_{30}}$ respectively, as shown in algorithm \ref{alg:convergence}.
    
\begin{figure}[t]
    \centering
    \begin{subfigure}[b]{0.55\linewidth}
        \includegraphics[width=1.0\linewidth]{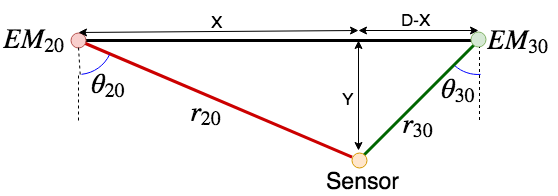}
        \caption{\label{fig:goemetry}}
    \end{subfigure}
    \hfill
    \begin{subfigure}[b]{0.44\linewidth}
        \includegraphics[width=1.0\linewidth]{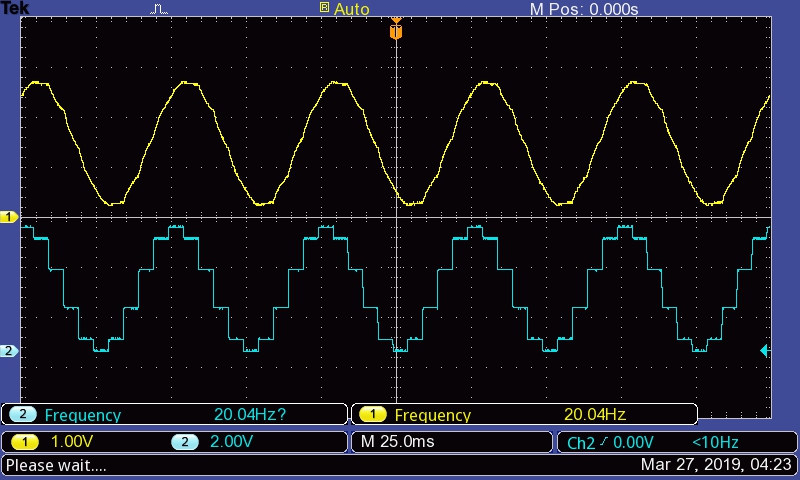}
        \caption{\label{fig:sinewave}}
    \end{subfigure}
    \caption{(a) Geometry of our setup. $EM_{20}$ and $EM_{30}$ are the centers of the electromagnets generating 20Hz and 30Hz sine waves respectively. (X,Y) is the position of our sensor considering $EM_{20}$ as origin, and, (b) 20Hz sine wave produced by the tone generating circuit. The Blue waveform is the raw output of the DAC. The Yellow waveform is after low pass filtering.}
\end{figure}

    The \textit{K} term in the algorithm is the constant determined by the physical properties of the electromagnet. As both our electromagnets are the same, we can use the same K for both of them. The \textit{D} term is also a constant as it is the distance between the centers of the 2 electromagnets as shown in Figure \ref{fig:goemetry}. The distance between our sensor and the 20Hz electromagnet and the 30Hz electromagnet is represented by $r_{20}$ and $r_{30}$ respectively and the angle is represented by $\theta_{20}$ and $\theta_{30}$. Initially, both $\theta_{20}$ and $\theta_{30}$ are set to 0. Values of $r_{20}$ and $r_{30}$ are calculated using these values of $\theta$ by using equation \ref{eq:mag_strength}. After obtaining the distance from the 2 electromagnets, relative position of the sensor represented by X and Y can be calculated by standard geometric equations. Once we obtain an initial value of X and Y, we can update our values of $\theta_{20}$ and $\theta_{30}$. 
    
    We conducted tests to see how fast the two $\theta s$ converge and observed that they converged in under 5 iterations in almost all cases. Thus N can be set to 5 without any loss in accuracy. Furthermore, after the initial convergence to the correct $\theta$ values, the next $\theta$ values can use the previous calculated $\theta$ values as an initial guess for faster convergence. We consider our update rate of 100Hz to be sufficient as it is very unlikely that a user's finger will move a large distance within tens of miliseconds. Consequently, the previous $\theta$ values will be very close to the present $\theta$ values.
    
    \subsection{Calibration}
    Before the user can begin using the system, a calibration procedure is needed. This step takes no more than a few seconds and requires the user to place his finger on a marked predefined position on the electromagnet bar while wearing the sensor. The calibration position on the bar was set to the midway point between the two electromagnets. During the calibration step, the system calculates the \textbf{K} term in equation \ref{eq:mag_strength} for the electromagnets. We have included this step to make our system compatible with any kind of magnet being driven by any amount of current. As \textbf{K} captures most of the magnets' parameters, as will be explained in Section \ref{sec:disc}, calibrating our system makes it magnet invariant. Additionally, constant field disturbances present in the environment are accounted for to a certain degree in the calibration step.
    
    \subsection{Localization Algorithm}
    Our algorithm consists of the following steps:
    \begin{enumerate}
        \item Read magnetic sensor value and add it to input buffer
        \item For all 3 axes, apply band-pass filtering and do an FFT on the filtered value.
        \item From the FFT calculate $\mathbf{H_{20}}$ and $\mathbf{H_{30}}$
        \item Apply algorithm \ref{alg:convergence} to obtain the coordinates of the sensor
    \end{enumerate}
    The input buffer can store up to 50 sensor readings. Thus each new sample is combined with 49 previous samples to compute the FFT. We have incorporated a combination of multithreaded and single threaded segments in our algorithm. The data collection and finding positions are done in serial while the signal conditioning and FFT is done in parallel for each axis. All our code is implemented in python using Numpy and Scipy libraries. We have also used the win32api library to control the mouse pointer from our python program.
    
    Implementing the algorithm on a PC or laptop is straightforward. Taking it a step further we implemented the tracking algorithm on the embedded micro-controller itself. It had a hardware Floating Point Unit(FPU) to speed up floating point calculations. The signal acquisition, filtering, buffering, FFT calculation and convergence algorithm were all implemented on the embedded processor. Due to lack of memory and compute power a smaller buffer of 20 samples was used, this did not lead to significant tracking accuracy loss. The system was timed to run the entire algorithm in under 5ms. This indicated even at a sampling rate of 200Hz the system will be able to process and track the user's finger. Processing on the edge like so reduces the burden of the central computer having to dedicate compute cycles for running the algorithm. This computation might be significant if multi-finger tracking is implemented. At a 100Hz update rate performing the computation on chip consumed around 2mA current on average. The battery we used was rated at 20mAh which means the system can be used for around 10 hours continuously before needing to be recharged.

\begin{table}
    \centering
    \begin{tabular}{|c|c|}
        \hline
         \textbf{Material} & \textbf{Mean Error(cm)} \\
         \hline\hline
         Metallic Tabletop & 0.21 $\pm$ 0.12\\
         \hline
         Wood & 0.20 $\pm$ 0.13\\
         \hline
         Acrylic & 0.22 $\pm$ 0.11\\
         \hline
    \end{tabular}
    \caption{Table showing average localization errors on different surfaces over the 25 points}
    \label{table:error}
\end{table}

\section{Hardware}
The MagSurface hardware consists of two components:
\begin{itemize}
    \item An electromagnet bar with two electromagnets, driven by an AC signal source
    \item A wireless wearable device containing the magnetometer sensor for location tracking and a micro-controller unit (MCU) with a BLE radio for data transmission
\end{itemize}

    \subsection{Electromagnet Bar}
    The electromagnet bar contains 2 electromagnets with their magnetic axis exactly parallel to each other. This reduces the computations involved when calculating the position of the finger. The 2 electromagnet centers have to be kept level and separated by a fixed distance. This is achieved by mounting them on a 3D printed enclosure.

    The electromagnets generate sinusoidal magnetic waves. AC fields as opposed to DC magnetic fields are used as it is much easier to reduce interference due to stray field with a tuned filter. Another advantage is that the earth's substantial magnetic field will not interfere with our measurements as that is a DC value which can be filtered out. A class D audio amplifier was used to drive the electromagnets; it provides stereo channel audio output, making it ideal for driving the 2 electromagnets. Furthermore, a gain control for tuning the current was also incorporated. To generate the tone, we employed an Atmega328P and a dual channel Digital to Analog Converter (DAC) to generate sinusoidal wave-forms at frequencies of 20Hz and 30Hz as shown in figure \ref{fig:sinewave}. The DAC produces discrete level sine waves, which are smoothed with a low pass filter.
    
    
    \subsection{Finger Tracking Wearable}
    To sense the magnetic fields, we built a small and lightweight wireless wearable device (See Figure \ref{fig:device_img}). The device was built on a flexible PCB and consists of three functional blocks: wireless MCU, sensing unit, and antenna. For the wireless MCU, CC2640 from Texas Instruments was used due to its low power consumption and small package size. As for the sensing subsystem, a MPU-9250 module\cite{mpu9250} from InvenSense was used. The MPU-9250 contains a 6-axis intertial motion unit (IMU) and a 3-axis magnetometer. For this specific application, only 3-axis magnetometer sensor was activated. Lastly, in the antenna block, we used a 2.45 GHz chip antenna and an impedance-matched balun filter from Johanson Technology. Touch on surfaces is detected by a pressure sensor connected to the MCU's digital input pin.

\begin{figure}[t]
    \centering
    \begin{subfigure}[b]{0.45\linewidth}
        \includegraphics[width=1.0\linewidth]{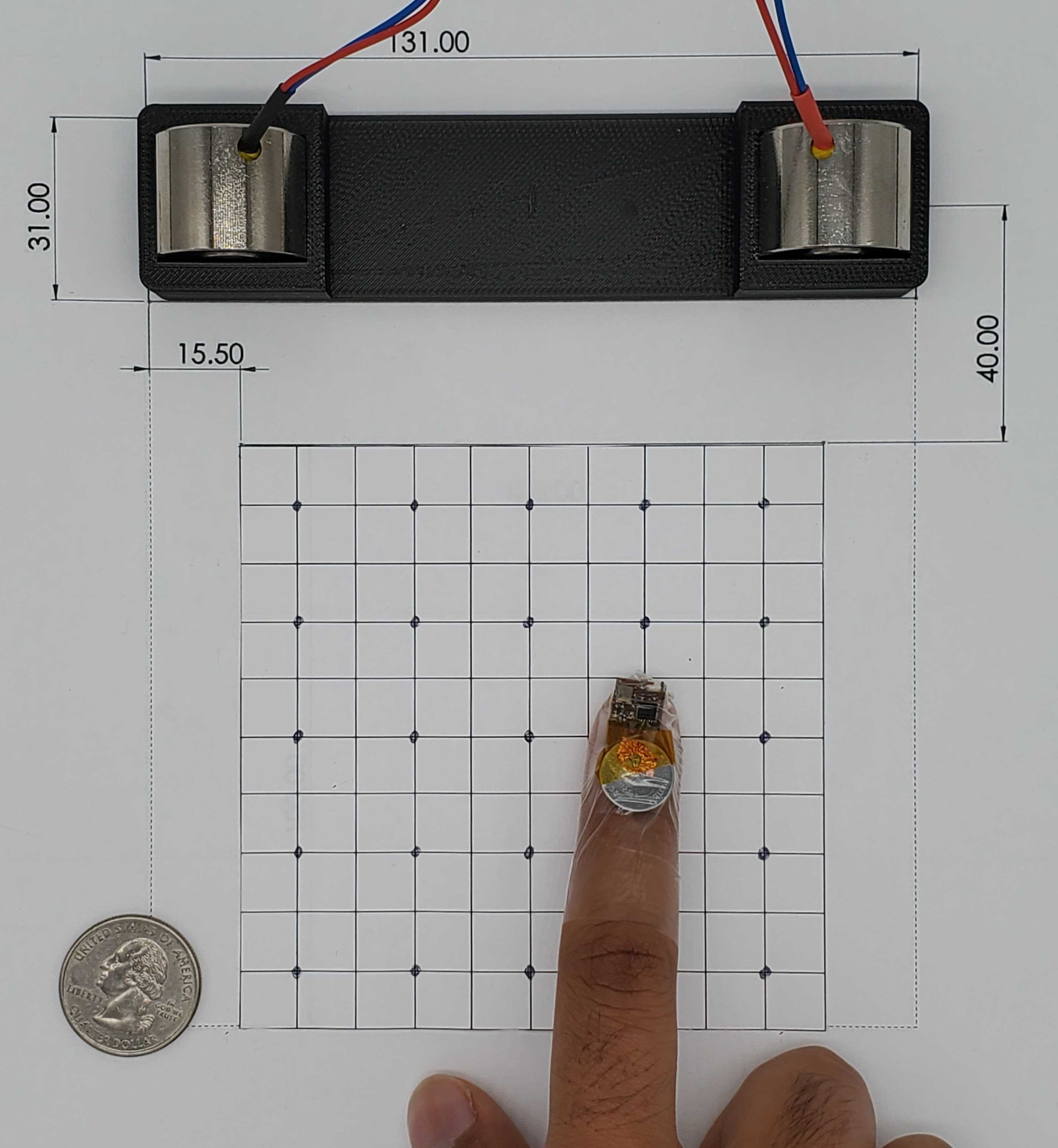}
        \caption{\label{fig:testbench}}
    \end{subfigure}
    \hfill
    \begin{subfigure}[b]{0.54\linewidth}
        \includegraphics[width=1.0\linewidth]{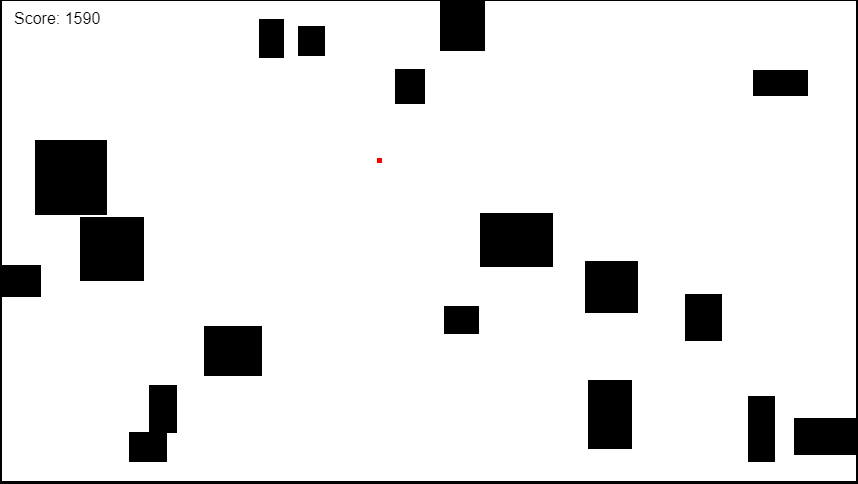}
        \begin{minipage}{0.1cm}
            \vspace{1.5cm}
        \end{minipage}
        \caption{\label{fig:Game}}
    \end{subfigure}
    \caption{(a) The test bench used for testing the accuracy of the system on various surfaces. 25 points on a grid of 10x10 $cm^2$ dimension was used to evaluate the system at different sensor positions. Measurements shown on the image are in millimeter, and, (b) Screenshot from the game the participants were asked to play. The Red dot was controlled by finger movements to dodge the black blocks.}
\end{figure}


\section{Evaluation and Results}
Localization accuracy of the system was evaluated and a user study to understand the acceptance of the system among users was conducted.

\begin{figure}[t]
    \centering
    \begin{subfigure}[b]{0.33\linewidth}
        \includegraphics[width=1\linewidth]{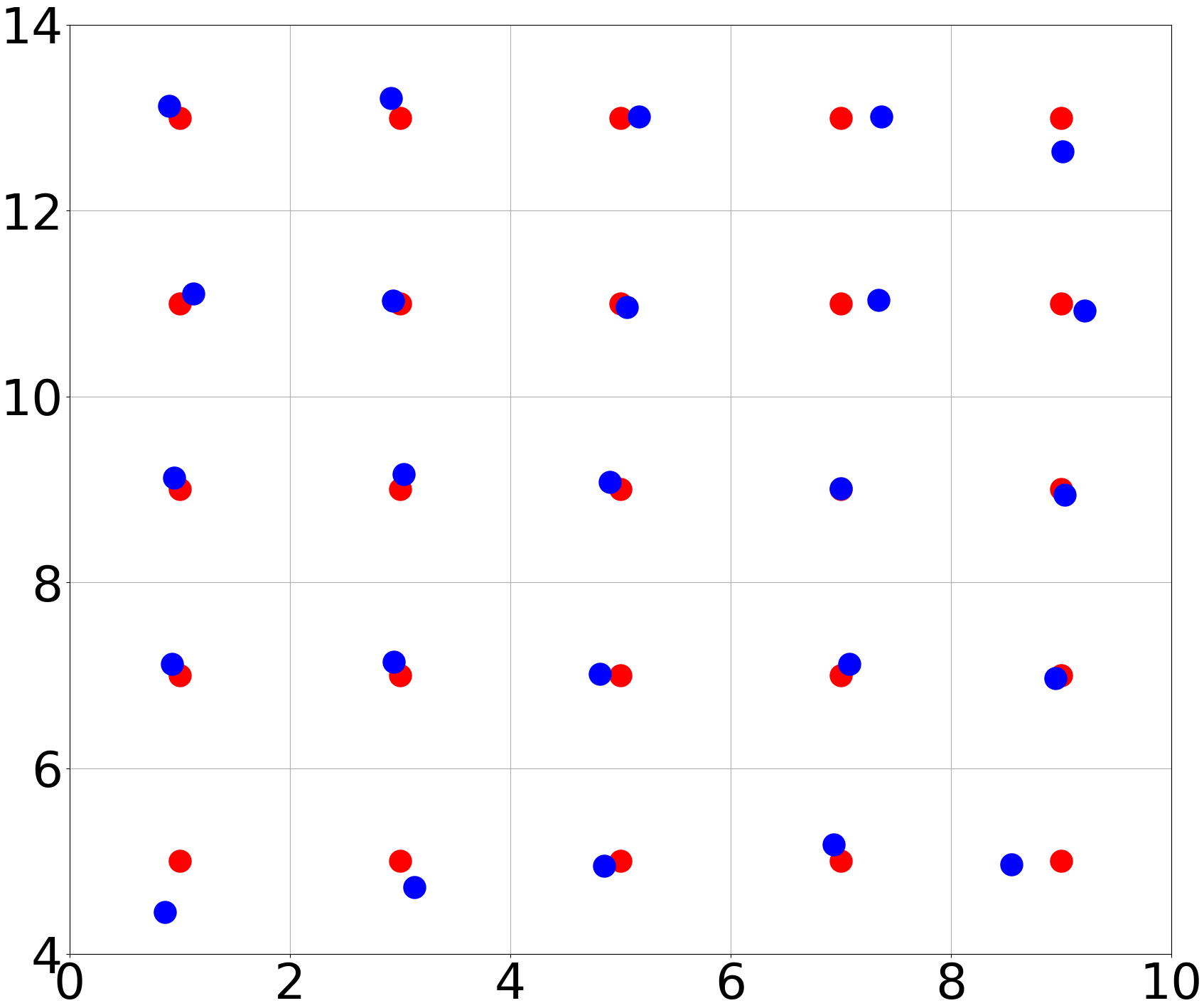}
        \caption{}
    \end{subfigure}
    \begin{subfigure}[b]{0.33\linewidth}
        \includegraphics[width=1\linewidth]{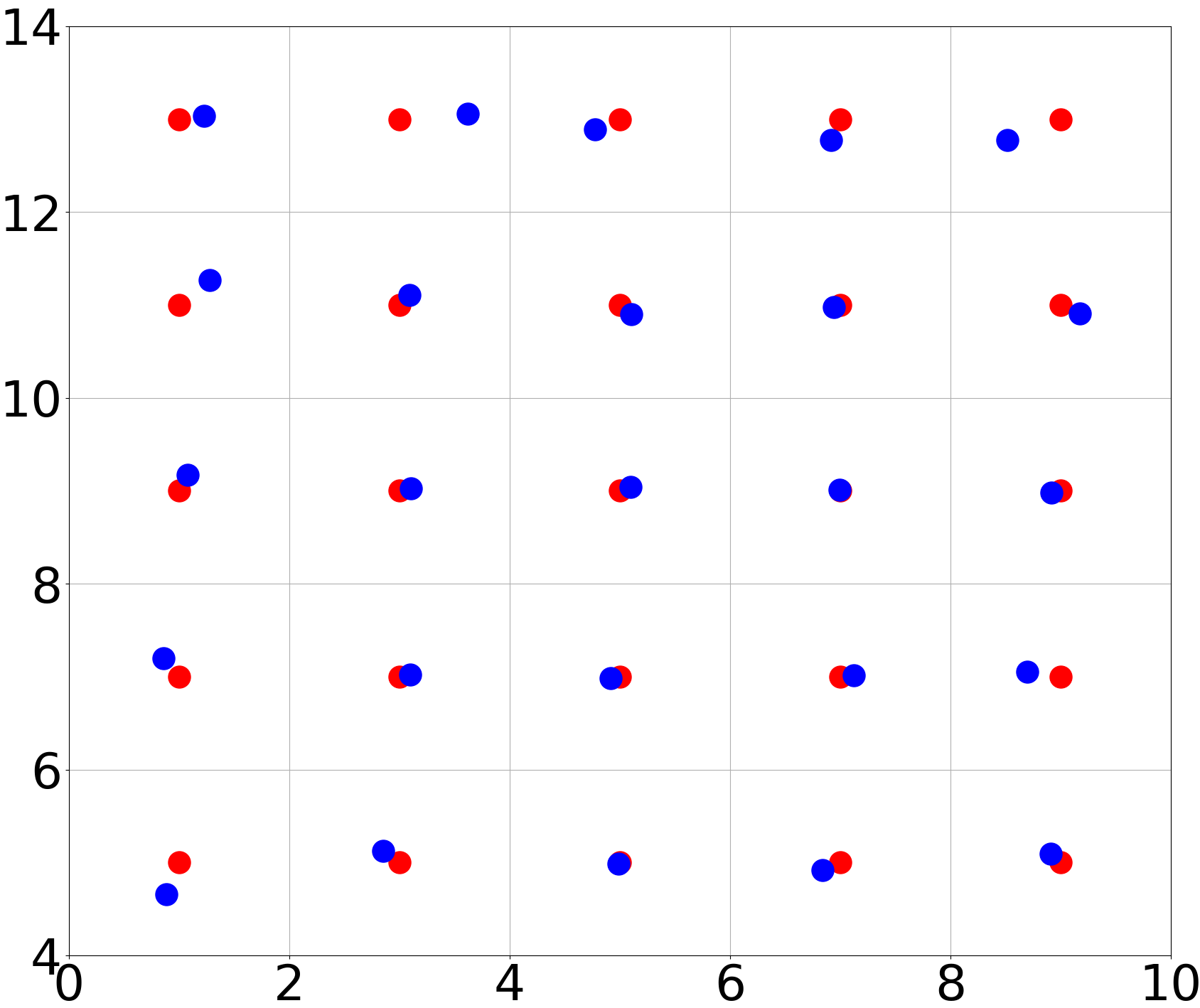}
        \caption{}
    \end{subfigure}
    \begin{subfigure}[b]{0.33\linewidth}
        \includegraphics[width=1\linewidth]{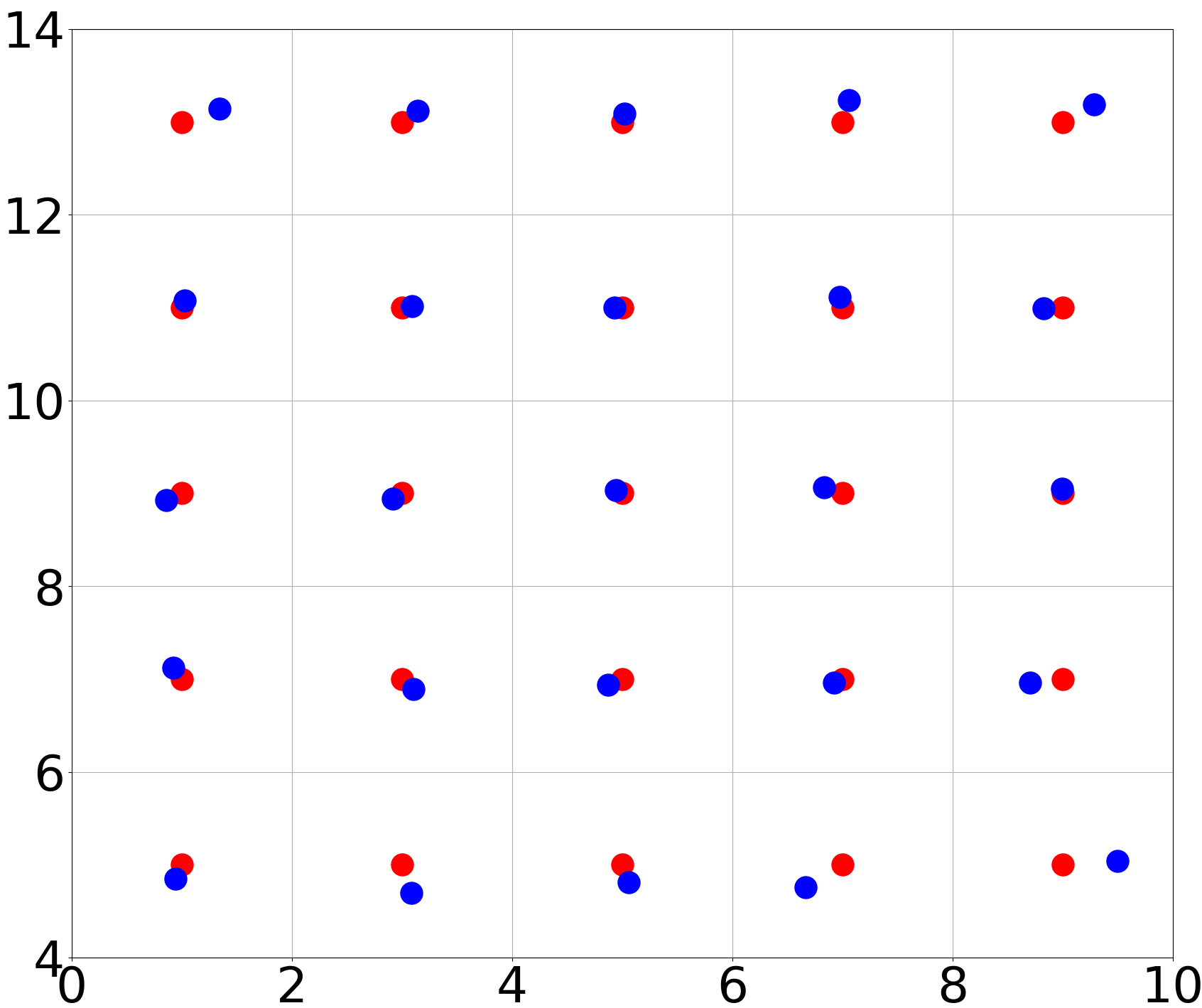}
        \caption{}
    \end{subfigure}
    \caption{Scatter plots showing the actual position (red) and the system determined position (blue) on different surfaces namely (a)Metal, (b)Wood, and (c)Acrylic. The axes measurements are in cm. The scatter plots represent the evaluation grid shown in Figure \ref{fig:testbench}.}
    \label{fig:scattererror}
\end{figure}

    \subsection{Localization Accuracy}
    The accuracy of tracking was measured on a 5x5 grid of points (Figure \ref{fig:testbench}). The relative locations of the grid points with respect to the electromagnetic bar were predefined. When a user places a finger on these predefined points while wearing the device, the tracking algorithm predicts the location of the finger. The accuracy was computed by obtaining the mean absolute error (MAE) between the predicted location and the true location across the 25 center points of the 5x5 grid (Table \ref{table:error}). This approach is similar to the one used by \cite{4421009}. We also examined the accuracy of the system on three different surfaces namely acrylic, wood and metal as these three surfaces are widely used in our daily lives. The scatter plots in figure \ref{fig:scattererror} depict the location of the 25 grid points (ground truth, in red) and the inferred location of the finger for each one of the points (in blue) for each one of the 3 surfaces tested.
    
    \subsection{Qualitative User Experience Study}
    A user study with a convenience sample of 12 participants (9 male, 3 female) aged 19 to 26 years-old was conducted to gain insights into the overall user experience of the system. In the study, participants wore the tracking device on their index finger and used it like a pointing device to play a game. The game consisted of moving a red dot and dodging moving blocks on the computer screen. Figure \ref{fig:Game} shows a screenshot of the game. Over time, the number of blocks and the speed at which they moved increased. In other words the difficulty increased as time passed. Participants were given 3 tries. After that the participants were asked to fill a survey which asked about the user experience of using the system.
    
    Overall, participants reacted positively to the experience (\textit{"This seems to not be tiring or hard to use for extended amount of time", "its nice to use my own finger to make movement, mouse can sometimes cause pain"}). A couple of participants commented on latency (\textit{"sometimes there is delay between of the movement between finger and controller", "I would love to suggest fixing the lagging so its more responsive and accurate"}). Finally, some participants also noticed a small amount of cursor jitter (\textit{"it would be good if the movement also stops when i am idle", "at moving the jitter is negligible but at slower speed or standstill, the jitter becomes increasingly noticeable"}). It was observed that this jitter was maximum at distances far away from the magnets as well as distances closer to the magnets. This corresponds to the maximum error regions that we observed during our evaluation of the accuracy of the system.

\section{Discussion}

\label{sec:disc}
    \subsection{Results}
    An interesting observation from the error plots is that the localization error is more pronounced at points further away and also very close to the electromagnets. Error at distant points is to be expected as the magnetic fields decay very rapidly with distance. For distances very close to the magnet, the error can be explained by the characteristics of our model. As stated earlier, the magnetic equations consider the electromagnets as small magnetic dipoles. But at distances very close to the electromagnets, the dimension of the electromagnets themselves are comparable to the distance from the magnet to the sensor, which results in inaccuracies. Another observation is that the error on different surfaces remained more or less same which indicated that the calibration step might be compensating for differences in environmental conditions.
    
    \subsection{User Feedback}
    The most prominent feedback was about the shakiness of the cursor when the participant kept their finger idle. Due to small variations in the output of the localization algorithm, in each iteration the new position changed slightly. This in turn caused the cursor on the screen to shake. A fix to this would be to introduce a dead-zone in the sensor movement, i.e., if the movement is small, ignore it completely. Another common feedback was to "fix the lagging so the system is more responsive". We use a sampling rate of 100Hz for the magnetometer. But the samples go to a FiFo which is 50 samples long for calculating the FFT. So there is a small latency in the movement of the finger being translated to a cursor movement. One way to address this issue would be to increase the sampling rate. There are magnetometers that can sample at 800Hz which would make the system much more responsive to position changes. Reducing the buffer size to 20 for our on device processing did improve the response rate by a good margin.

    \subsection{Custom Electromagnets}
    For our electromagnets, we used 2 off-the-shelf electromagnets whose exact specifications we could not confirm. This was possible because, as previously described, our calibration step makes it possible for our system to be electromagnet invariant. We compute the value for the magnetic field strength \textbf{K} used in equation \ref{eq:K}. \textbf{K} is proportional to 4 factors:
    
    \begin{equation}
    \textbf{K} \propto \mathbf{\mu} \times \textbf{N} \times \textbf{I} \times \textbf{A}
    \label{eq:K}
    \end{equation}
    \newline
    $\mathbf{\mu}$ is the magnetic permeability of the core material, \textbf{N} is the number of turns of the coil, \textbf{I} is the current flowing through the coil and \textbf{A} is the cross sectional area of the coil.
    
    In designing custom electromagnets, we would choose to use a core material with larger relative magnetic permeability. This would allow us to reduce the cross sectional area and number of turns, thereby reducing the size of the electromagnet without reducing the magnetic field strength of the generated fields. It would also enable us to reduce the current flowing through the coil thus making the system more power efficient.
    
    \subsection{Enlarging Tracking Area}
    The operation of MagSurface is by no means restricted to the 10x10 \(cm^2\) rectangular area used to evaluate it. There are 2 ways to increase the size of the finger tracking space: (1) make the generated magnetic field stronger, or (2) make the sensors more sensitive.
    
    The magnetic field can be strengthened by raising K, which can be achieved by increasing any of the 4 factors present in equation \ref{eq:K}. We would not want to increase A or N as that would make the system bulky and reduce portability. Alternatively, we could use better core materials or drive more current. These approaches come with their own limitations. Using a better core material will make the system expensive, and increasing current would make the system energy inefficient and also cause heating issues. We have to consider all these trade-offs when designing the system to balance both the pros and the cons.
    
    Using better sensors would be another option. The sensor that we are currently using has a resolution of 0.6$\mu$T. There are sensors which have a resolution of 0.01$\mu$T. But that high level of sensitivity also comes with trade-offs, either they have a low sampling rate, or they are very expensive, or they have a large physical size. For our purpose, the system requires high sampling rate for low latency tracking and small physical size for portability and ease of use.
    
    There is still a limit to how big of an interaction space can be used. As magnetic fields are inversely proportional to the cube of the distance, increasing the magnetic strength by 8 would only lead to a sensing distance increase of 2. So there is a very sharp diminishing return.
    
    \subsection{Richer Interactions}
    In this work, we demonstrate the viability of MagSurface with just one magnetometer, placed on the finger. Since the magnetometer does not affect the magnetic field being generated, an interactive experience leveraging multiple fingers, and thus multiple magnetometers, could be designed with our proposed approach. In future work, we would like to investigate the space of gestural interactions using more than one finger, e.g., zoom, scroll and multi-finger swipe. Additionally, multi-user interaction in the same space is another area that can be explored.
    
    The number of sensors that can be added would be limited by the BLE data rate. Each sensor would have a rate at which it would send its position. The maximum number of sensors \textbf{n} theoretically supported would be
    \begin{equation}
    \textbf{n} = \frac{max\ BLE\ throughput}{Device\ update\ rate \times Data\ per\ update}
    \label{eq:devices}
    \end{equation}

\section{Usage Scenarios}
We envision the use of MagSurface both as a way to extend the input space of existing devices and as replacement to existing touchscreens and touchpads in specific scenarios where these technologies might present shortcomings.

    \subsubsection*{\textbf{Extended Interactive Surface}}
    By incorporating MagSurface into computing devices such as laptops and tablets, the touch-interactive surface of these devices could be expanded beyond their built-in touchpad or touchscreen. Furthermore as the wearable sensing device can sit on top of a finger (i.e., on the fingernail), or placed on a ring, it would not hinder other forms of input like typing on the keyboard or using a mouse. This usage scenario is similar to the one laid out by Mistry and Maes \cite{Mistry:2011:MCM:1979742.1979715}.
    
    \subsubsection*{\textbf{Smart Furniture}}
    As smart rooms become more common, MagSurface provides an opportunity to embedded touch input into furniture without modifying its design and aesthetic. As an example, MagSurface's electromagnets could be placed within a sofa arm, turning the arm into a tracking surface. By wearing the wireless finger tracking device on a ring, users could control a smart TV, sound system or IoT devices without the need for an additional device (e.g., remote control with a trackpad). This scenario is possible because MagSurface permits finger tracking even on surfaces that are corrugated or covered by fabric, i.e., not perfectly flat.
    
    \subsubsection*{\textbf{Wet Lab and Surgery Room}}
    In a wet lab or surgery room, where scientists and doctors might handle hazardous materials or be in contact with contagious substances (e.g., blood), maintaining a sterile environment is paramount. For this reason, hands must be regularly de-germed and the utilization of gloves is a strict requirement. Gloves should be taken off during computer use however, which is highly disruptive. Computer hardware, especially keyboards and touch surfaces, can be contaminated with microorganisms when touched by contaminated hands. By wearing MagSurface's wireless finger tracking device, it would be possible to let doctors and scientists interact with a computer via touch while wearing gloves on a surface which can be heavily disinfected with chemicals that might not be suitable for existing touchscreen electronics.
    
    \subsubsection*{\textbf{COVID-19}}
    The COVID-19 virus is very infectious and is known to spread indirectly via contact with surfaces. In many places such as airport check self check in counters, grocery self checkout, etc. people have to use a touchscreen. In this situation strong disinfectants need to be used on public touchscreens to prevent the spread of the virus. Using such strong chemicals can cause damage to traditional touchscreens, replacement of which can be a costly affair. Our approach lets any surface become a touchscreen, thus enabling regular tabletops of even replaceable and disposable plastic or paper sheets to become touch surfaces. Additionally the wearable part can be embedded inside a sealed plastic mould which can be clipped onto the finger. Thus the wearable device can also be disinfected after use.

\section{Conclusion}
This paper presents an approach that leverages magnetic fields to enable finger tracking on any flat surface. We demonstrated an implementation of our approach with MagSurface, a system comprised of an electromagnet bar and a wearable band-aid. We performed an evaluation of the system on 3 surface materials, metallic, wood and acrylic, in which we showed that our approach produces localization errors not greater than 0.25cm. Despite opportunities for improvements, our method proved viable as a way to bring finger tracking to settings where a perfectly smooth, clean and dry sensing surface cannot be expected.

\bibliographystyle{ACM-Reference-Format}

\begin{thebibliography}{00}


\ifx \showCODEN    \undefined \def \showCODEN     #1{\unskip}     \fi
\ifx \showDOI      \undefined \def \showDOI       #1{{\tt DOI:}\penalty0{#1}\ }
  \fi
\ifx \showISBNx    \undefined \def \showISBNx     #1{\unskip}     \fi
\ifx \showISBNxiii \undefined \def \showISBNxiii  #1{\unskip}     \fi
\ifx \showISSN     \undefined \def \showISSN      #1{\unskip}     \fi
\ifx \showLCCN     \undefined \def \showLCCN      #1{\unskip}     \fi
\ifx \shownote     \undefined \def \shownote      #1{#1}          \fi
\ifx \showarticletitle \undefined \def \showarticletitle #1{#1}   \fi
\ifx \showURL      \undefined \def \showURL       #1{#1}          \fi

\bibitem{evomouse}
 2010.
\newblock EvoMouse.
\newblock   (2010).
\newblock
\showURL{%
\url{https://www.celluon.com/evomouse/}}


\bibitem{leap}
 2010.
\newblock Leap Motion.
\newblock   (2010).
\newblock
\showURL{%
\url{https://www.leapmotion.com/}}


\bibitem{mpu9250}
 2015.
\newblock InvenSense, MPU-9250 Nine-Axis MEMS MotionTracking Device.
\newblock   (2015).
\newblock
\showURL{%
\url{http://www.invensense.com/-products/motion-tracking/9-axis/mpu-9250/}}


\bibitem{Ashbrook:2011:NSE:1978942.1979238}
{Daniel Ashbrook}, {Patrick Baudisch}, {and} {Sean White}. 2011.
\newblock \showarticletitle{Nenya: Subtle and Eyes-free Mobile Input with a
  Magnetically-tracked Finger Ring}. In {\em Proceedings of the SIGCHI
  Conference on Human Factors in Computing Systems} {\em (CHI '11)}. ACM, New
  York, NY, USA, 2043--2046.
\newblock
\showISBNx{978-1-4503-0228-9}
\showDOI{%
\url{http://dx.doi.org/10.1145/1978942.1979238}}


\bibitem{Chen:2013:UIU:2501988.2502035}
{Ke-Yu Chen}, {Kent Lyons}, {Sean White}, {and} {Shwetak Patel}. 2013.
\newblock \showarticletitle{uTrack: 3D Input Using Two Magnetic Sensors}. In
  {\em Proceedings of the 26th Annual ACM Symposium on User Interface Software
  and Technology} {\em (UIST '13)}. ACM, New York, NY, USA, 237--244.
\newblock
\showISBNx{978-1-4503-2268-3}
\showDOI{%
\url{http://dx.doi.org/10.1145/2501988.2502035}}


\bibitem{Chen:2016:FTP:2858036.2858125}
{Ke-Yu Chen}, {Shwetak~N. Patel}, {and} {Sean Keller}. 2016.
\newblock \showarticletitle{Finexus: Tracking Precise Motions of Multiple
  Fingertips Using Magnetic Sensing}. In {\em Proceedings of the 2016 CHI
  Conference on Human Factors in Computing Systems} {\em (CHI '16)}. ACM, New
  York, NY, USA, 1504--1514.
\newblock
\showISBNx{978-1-4503-3362-7}
\showDOI{%
\url{http://dx.doi.org/10.1145/2858036.2858125}}


\bibitem{dai20176}
{Houde Dai}, {Shuang Song}, {Xianping Zeng}, {Shjian Su}, {Mingqiang Lin},
  {and} {Max Q-H Meng}. 2017.
\newblock \showarticletitle{6-D electromagnetic tracking approach using
  uniaxial transmitting coil and tri-axial magneto-resistive sensor}.
\newblock {\em IEEE Sensors Journal\/} {18}, 3 (2017), 1178--1186.
\newblock


\bibitem{de2010function}
{Isabelo de~los Reyes}, {Nathanael Roberton}, {Brian Calvery}, {Timothy~JE
  Turner}, {Adrian Chandley}, {Daniel Makoski}, {Paul Henderson}, {Egor
  Nikitin}, {Tarek Elabbady}, {Phillip Joe}, {and} {others}. 2010.
\newblock Function-oriented user interface.
\newblock   (June~1 2010).
\newblock
\newblock
\shownote{US Patent 7,730,425.}


\bibitem{Echtler:2009:IFE:1731903.1731909}
{Florian Echtler}, {Andreas Dippon}, {Marcus T\"{o}nnis}, {and} {Gudrun
  Klinker}. 2009.
\newblock \showarticletitle{Inverted FTIR: Easy Multitouch Sensing for
  Flatscreens}. In {\em Proceedings of the ACM International Conference on
  Interactive Tabletops and Surfaces} {\em (ITS '09)}. ACM, New York, NY, USA,
  29--32.
\newblock
\showISBNx{978-1-60558-733-2}
\showDOI{%
\url{http://dx.doi.org/10.1145/1731903.1731909}}


\bibitem{epps1986comparison}
{Brian~W Epps}. 1986.
\newblock \showarticletitle{Comparison of six cursor control devices based on
  Fitts' law models}. In {\em Proceedings of the Human Factors Society Annual
  Meeting}, Vol.~30. SAGE Publications Sage CA: Los Angeles, CA, 327--331.
\newblock


\bibitem{ge2009novel}
{Xin Ge}, {Dakun Lai}, {Xiaomei Wu}, {and} {Zuxiang Fang}. 2009.
\newblock \showarticletitle{A novel non-model-based 6-DOF electromagnetic
  tracking method using non-iterative algorithm}. In {\em 2009 Annual
  International Conference of the IEEE Engineering in Medicine and Biology
  Society}. IEEE, 5144--5117.
\newblock


\bibitem{4421009}
{Xinying Han}, {Hiroaki Seki}, {Yoshitsugu Kamiya}, {and} {Masatoshi Hikizub}.
  2007.
\newblock \showarticletitle{Wearable handwriting input device using magnetic
  field}. In {\em SICE Annual Conference 2007}. 365--368.
\newblock
\showDOI{%
\url{http://dx.doi.org/10.1109/SICE.2007.4421009}}


\bibitem{Harrison:2011:OWM:2047196.2047255}
{Chris Harrison}, {Hrvoje Benko}, {and} {Andrew~D. Wilson}. 2011.
\newblock \showarticletitle{OmniTouch: Wearable Multitouch Interaction
  Everywhere}. In {\em Proceedings of the 24th Annual ACM Symposium on User
  Interface Software and Technology} {\em (UIST '11)}. ACM, New York, NY, USA,
  441--450.
\newblock
\showISBNx{978-1-4503-0716-1}
\showDOI{%
\url{http://dx.doi.org/10.1145/2047196.2047255}}


\bibitem{Harrison:2009:AWH:1622176.1622199}
{Chris Harrison} {and} {Scott~E. Hudson}. 2009.
\newblock \showarticletitle{Abracadabra: Wireless, High-precision, and
  Unpowered Finger Input for Very Small Mobile Devices}. In {\em Proceedings of
  the 22Nd Annual ACM Symposium on User Interface Software and Technology} {\em
  (UIST '09)}. ACM, New York, NY, USA, 121--124.
\newblock
\showISBNx{978-1-60558-745-5}
\showDOI{%
\url{http://dx.doi.org/10.1145/1622176.1622199}}


\bibitem{Hofer:2009:FFM:1517664.1517730}
{Ramon Hofer}, {Daniel Naeff}, {and} {Andreas Kunz}. 2009.
\newblock \showarticletitle{FLATIR: FTIR Multi-touch Detection on a Discrete
  Distributed Sensor Array}. In {\em Proceedings of the 3rd International
  Conference on Tangible and Embedded Interaction} {\em (TEI '09)}. ACM, New
  York, NY, USA, 317--322.
\newblock
\showISBNx{978-1-60558-493-5}
\showDOI{%
\url{http://dx.doi.org/10.1145/1517664.1517730}}


\bibitem{hutchins1985direct}
{Edwin~L Hutchins}, {James~D Hollan}, {and} {Donald~A Norman}. 1985.
\newblock \showarticletitle{Direct manipulation interfaces}.
\newblock {\em Human-computer interaction\/} {1}, 4 (1985), 311--338.
\newblock


\bibitem{Kim:2018:ULA:3274783.3274848}
{Hyosu Kim}, {Anish Byanjankar}, {Yunxin Liu}, {Yuanchao Shu}, {and} {Insik
  Shin}. 2018.
\newblock \showarticletitle{UbiTap: Leveraging Acoustic Dispersion for
  Ubiquitous Touch Interface on Solid Surfaces}. In {\em Proceedings of the
  16th ACM Conference on Embedded Networked Sensor Systems} {\em (SenSys '18)}.
  ACM, New York, NY, USA, 211--223.
\newblock
\showISBNx{978-1-4503-5952-8}
\showDOI{%
\url{http://dx.doi.org/10.1145/3274783.3274848}}


\bibitem{Liang:2012:GAS:2380116.2380157}
{Rong-Hao Liang}, {Kai-Yin Cheng}, {Chao-Huai Su}, {Chien-Ting Weng}, {Bing-Yu
  Chen}, {and} {De-Nian Yang}. 2012.
\newblock \showarticletitle{GaussSense: Attachable Stylus Sensing Using
  Magnetic Sensor Grid}. In {\em Proceedings of the 25th Annual ACM Symposium
  on User Interface Software and Technology} {\em (UIST '12)}. ACM, New York,
  NY, USA, 319--326.
\newblock
\showISBNx{978-1-4503-1580-7}
\showDOI{%
\url{http://dx.doi.org/10.1145/2380116.2380157}}


\bibitem{7964907}
{J. {Liu}}, {Y. {Chen}}, {M. {Gruteser}}, {and} {Y. {Wang}}. 2017.
\newblock \showarticletitle{VibSense: Sensing Touches on Ubiquitous Surfaces
  through Vibration}. In {\em 2017 14th Annual IEEE International Conference on
  Sensing, Communication, and Networking (SECON)}. 1--9.
\newblock
\showISSN{2155-5494}
\showDOI{%
\url{http://dx.doi.org/10.1109/SAHCN.2017.7964907}}


\bibitem{Mistry:2011:MCM:1979742.1979715}
{Pranav Mistry} {and} {Pattie Maes}. 2011.
\newblock \showarticletitle{Mouseless: A Computer Mouse As Small As Invisible}.
  In {\em CHI '11 Extended Abstracts on Human Factors in Computing Systems}
  {\em (CHI EA '11)}. ACM, New York, NY, USA, 1099--1104.
\newblock
\showISBNx{978-1-4503-0268-5}
\showDOI{%
\url{http://dx.doi.org/10.1145/1979742.1979715}}


\bibitem{ono2013touch}
{Makoto Ono}, {Buntarou Shizuki}, {and} {Jiro Tanaka}. 2013.
\newblock \showarticletitle{Touch \& activate: adding interactivity to existing
  objects using active acoustic sensing}. In {\em Proceedings of the 26th
  annual ACM symposium on User interface software and technology}. ACM, 31--40.
\newblock


\bibitem{poupyrev2016project}
{Ivan Poupyrev}, {Nan-Wei Gong}, {Shiho Fukuhara}, {Mustafa~Emre Karagozler},
  {Carsten Schwesig}, {and} {Karen~E Robinson}. 2016.
\newblock \showarticletitle{Project Jacquard: interactive digital textiles at
  scale}. In {\em Proceedings of the 2016 CHI Conference on Human Factors in
  Computing Systems}. ACM, 4216--4227.
\newblock


\bibitem{4102227}
{F.~H. {Raab}}, {E.~B. {Blood}}, {T.~O. {Steiner}}, {and} {H.~R. {Jones}}.
  1979.
\newblock \showarticletitle{Magnetic Position and Orientation Tracking System}.
\newblock {\it IEEE Trans. Aerospace Electron. Systems} {AES-15}, 5 (Sep.
  1979), 709--718.
\newblock
\showISSN{0018-9251}
\showDOI{%
\url{http://dx.doi.org/10.1109/TAES.1979.308860}}


\bibitem{Rekimoto:2002:SIF:503376.503397}
{Jun Rekimoto}. 2002.
\newblock \showarticletitle{SmartSkin: An Infrastructure for Freehand
  Manipulation on Interactive Surfaces}. In {\em Proceedings of the SIGCHI
  Conference on Human Factors in Computing Systems} {\em (CHI '02)}. ACM, New
  York, NY, USA, 113--120.
\newblock
\showISBNx{1-58113-453-3}
\showDOI{%
\url{http://dx.doi.org/10.1145/503376.503397}}


\bibitem{Rosenberg:2009:UIM:1576246.1531371}
{Ilya Rosenberg} {and} {Ken Perlin}. 2009.
\newblock \showarticletitle{The UnMousePad: An Interpolating Multi-touch
  Force-sensing Input Pad}. In {\em ACM SIGGRAPH 2009 Papers} {\em (SIGGRAPH
  '09)}. ACM, New York, NY, USA, Article 65, 9 pages.
\newblock
\showISBNx{978-1-60558-726-4}
\showDOI{%
\url{http://dx.doi.org/10.1145/1576246.1531371}}


\bibitem{Sato:2012:TET:2207676.2207743}
{Munehiko Sato}, {Ivan Poupyrev}, {and} {Chris Harrison}. 2012.
\newblock \showarticletitle{Touch{\'e}: Enhancing Touch Interaction on Humans,
  Screens, Liquids, and Everyday Objects}. In {\em Proceedings of the SIGCHI
  Conference on Human Factors in Computing Systems} {\em (CHI '12)}. ACM, New
  York, NY, USA, 483--492.
\newblock
\showISBNx{978-1-4503-1015-4}
\showDOI{%
\url{http://dx.doi.org/10.1145/2207676.2207743}}


\bibitem{5655110}
{A. {Sulaiman}}, {K. {Poletkin}}, {and} {A.~W.~H. {Khong}}. 2010.
\newblock \showarticletitle{Source Localization in the Presence of Dispersion
  for Next Generation Touch Interface}. In {\em 2010 International Conference
  on Cyberworlds}. 82--86.
\newblock
\showDOI{%
\url{http://dx.doi.org/10.1109/CW.2010.72}}


\bibitem{Sun:2018:VST:3241539.3241568}
{Ke Sun}, {Ting Zhao}, {Wei Wang}, {and} {Lei Xie}. 2018.
\newblock \showarticletitle{VSkin: Sensing Touch Gestures on Surfaces of Mobile
  Devices Using Acoustic Signals}. In {\em Proceedings of the 24th Annual
  International Conference on Mobile Computing and Networking} {\em (MobiCom
  '18)}. ACM, New York, NY, USA, 591--605.
\newblock
\showISBNx{978-1-4503-5903-0}
\showDOI{%
\url{http://dx.doi.org/10.1145/3241539.3241568}}


\bibitem{Takeoka:2010:ZIG:1936652.1936668}
{Yoshiki Takeoka}, {Takashi Miyaki}, {and} {Jun Rekimoto}. 2010.
\newblock \showarticletitle{Z-touch: An Infrastructure for 3D Gesture
  Interaction in the Proximity of Tabletop Surfaces}. In {\em ACM International
  Conference on Interactive Tabletops and Surfaces} {\em (ITS '10)}. ACM, New
  York, NY, USA, 91--94.
\newblock
\showISBNx{978-1-4503-0399-6}
\showDOI{%
\url{http://dx.doi.org/10.1145/1936652.1936668}}


\bibitem{Wilson:2004:TIT:1027933.1027946}
{Andrew~D. Wilson}. 2004.
\newblock \showarticletitle{TouchLight: An Imaging Touch Screen and Display for
  Gesture-based Interaction}. In {\em Proceedings of the 6th International
  Conference on Multimodal Interfaces} {\em (ICMI '04)}. ACM, New York, NY,
  USA, 69--76.
\newblock
\showISBNx{1-58113-995-0}
\showDOI{%
\url{http://dx.doi.org/10.1145/1027933.1027946}}


\bibitem{Wilson:2005:PCI:1095034.1095047}
{Andrew~D. Wilson}. 2005.
\newblock \showarticletitle{PlayAnywhere: A Compact Interactive Tabletop
  Projection-vision System}. In {\em Proceedings of the 18th Annual ACM
  Symposium on User Interface Software and Technology} {\em (UIST '05)}. ACM,
  New York, NY, USA, 83--92.
\newblock
\showISBNx{1-59593-271-2}
\showDOI{%
\url{http://dx.doi.org/10.1145/1095034.1095047}}


\bibitem{Xiao:2013:WRE:2470654.2466113}
{Robert Xiao}, {Chris Harrison}, {and} {Scott~E. Hudson}. 2013.
\newblock \showarticletitle{WorldKit: Rapid and Easy Creation of Ad-hoc
  Interactive Applications on Everyday Surfaces}. In {\em Proceedings of the
  SIGCHI Conference on Human Factors in Computing Systems} {\em (CHI '13)}.
  ACM, New York, NY, USA, 879--888.
\newblock
\showISBNx{978-1-4503-1899-0}
\showDOI{%
\url{http://dx.doi.org/10.1145/2470654.2466113}}


\bibitem{Yoo:2016:SEN:2858036.2858286}
{Chungkuk Yoo}, {Inseok Hwang}, {Eric Rozner}, {Yu Gu}, {and} {Robert~F.
  Dickerson}. 2016.
\newblock \showarticletitle{SymmetriSense: Enabling Near-Surface Interactivity
  on Glossy Surfaces Using a Single Commodity Smartphone}. In {\em Proceedings
  of the 2016 CHI Conference on Human Factors in Computing Systems} {\em (CHI
  '16)}. ACM, New York, NY, USA, 5126--5137.
\newblock
\showISBNx{978-1-4503-3362-7}
\showDOI{%
\url{http://dx.doi.org/10.1145/2858036.2858286}}


\bibitem{Zhang:2017:SCT:3120957.3090095}
{Cheng Zhang}, {Qiuyue Xue}, {Anandghan Waghmare}, {Sumeet Jain}, {Yiming Pu},
  {Sinan Hersek}, {Kent Lyons}, {Kenneth~A. Cunefare}, {Omer~T. Inan}, {and}
  {Gregory~D. Abowd}. 2017b.
\newblock \showarticletitle{SoundTrak: Continuous 3D Tracking of a Finger Using
  Active Acoustics}.
\newblock {\em Proc. ACM Interact. Mob. Wearable Ubiquitous Technol.\/} {1}, 2,
  Article 30 (June 2017), 25 pages.
\newblock
\showISSN{2474-9567}
\showDOI{%
\url{http://dx.doi.org/10.1145/3090095}}


\bibitem{Zhang:2017:ELT:3025453.3025842}
{Yang Zhang}, {Gierad Laput}, {and} {Chris Harrison}. 2017a.
\newblock \showarticletitle{Electrick: Low-Cost Touch Sensing Using Electric
  Field Tomography}. In {\em Proceedings of the 2017 CHI Conference on Human
  Factors in Computing Systems} {\em (CHI '17)}. ACM, New York, NY, USA, 1--14.
\newblock
\showISBNx{978-1-4503-4655-9}
\showDOI{%
\url{http://dx.doi.org/10.1145/3025453.3025842}}


\bibitem{Zhang:2016:SUB:2858036.2858082}
{Yang Zhang}, {Junhan Zhou}, {Gierad Laput}, {and} {Chris Harrison}. 2016.
\newblock \showarticletitle{SkinTrack: Using the Body As an Electrical
  Waveguide for Continuous Finger Tracking on the Skin}. In {\em Proceedings of
  the 2016 CHI Conference on Human Factors in Computing Systems} {\em (CHI
  '16)}. ACM, New York, NY, USA, 1491--1503.
\newblock
\showISBNx{978-1-4503-3362-7}
\showDOI{%
\url{http://dx.doi.org/10.1145/2858036.2858082}}


\end{thebibliography}

\end{document}